\documentclass[12pt]{article}
\usepackage{amssymb,amsmath,epsfig}
\allowdisplaybreaks

\begin{document}

\title{\bf Extended Black Hole Solutions in Rastall Theory of Gravity}
\author{M. Sharif$^1$ \thanks{msharif.math@pu.edu.pk}~
and M. Sallah$^{1,2}$ \thanks{malick.sallah@utg.edu.gm} \\
$^1$ Department of Mathematics and Statistics, The University of Lahore\\
1-KM Defence Road Lahore-54000, Pakistan.\\
$^2$ Department of Mathematics, The University of The Gambia,\\
Serrekunda, P.O. Box 3530, The Gambia.}
\date{}
\maketitle

\begin{abstract}
We utilize the gravitational decoupling via the extended geometric
deformation to extend the Schwarzschild vacuum solution to new black
holes in Rastall theory. By employing linear transformations that
deform both the temporal and radial coefficients of the metric, the
field equations with a dual matter source are successfully decoupled
into two sets. The first of these sets is described by the metric
for the vacuum Schwarzschild spacetime, while the second set
corresponds to the added extra source. Three extended solutions are
obtained using two restrictions on the metric potentials and extra
source, respectively. For selected values of the Rastall and
decoupling parameters, we study the impact of the fluctuation of
these parameters on the obtained models. We also investigate the
asymptotic flatness of the resulting spacetimes by analysis of the
metric coefficients. Finally, the nature of the additional source is
explored for each model, via analysis of the energy conditions. It
is found among other results that none of the obtained models
satisfy the energy conditions, while only the model corresponding to
the barotropic equation of state mimics an asymptotically flat
spacetime.
\end{abstract}

\section{Introduction}

The assumption behind the Rastall theory of gravity \cite{1} is that
the laws of conservation such as the conservation of mass and energy
can only be studied within the flat or weak-field domain of
spacetime. Recently, a novel idea has been put forth which
introduces the non-minimal interaction between matter and
gravitational fields as the source of the universe accelerating
momentum \cite{2}. Rastall argument allows one to add new elements
to the Einstein field equations by relaxing the requirement that the
covariant derivative of the energy-momentum tensor be zero. Indeed,
it has recently been demonstrated that in a curved spacetime, the
divergence of the energy-momentum tensor can be non-zero \cite{3}.
Numerous precise solutions have been found for this theory in both
astrophysical \cite{4}-\cite{8} and cosmological contexts
\cite{9}-\cite{13}.

By contrasting the thermodynamic quantities and features of black
holes in Rastall gravity with those found in general relativity
(GR), the Rastall hypothesis allows us to better appreciate the
quality of the connection between geometry and matter fields, which
is non-minimal. The Einstein field equations produce numerous
solutions that exhibit the geometric structure of spacetime. An
object known as a black hole (a region of spacetime with a
coordinate singularity given by an event horizon, beyond which
nothing escapes its strong gravitational pull) is the result of many
solutions of the Einstein field equations. Some well-known black
hole solutions are developed by Schwarzschild, Riessner-Nordstrom,
Kerr and Kerr-Newmann. Apart from the theoretical descriptions,
gravitational waves resulting from the merger of two black holes
have been observed by LIGO \cite{14}, and the event horizon
telescope has recently obtained the first-ever actual image of a
black hole \cite{15}. Black hole solutions including but not limited
to the static black hole \cite{16} and the rotating charged black
hole solutions \cite{17}, have been found in Rastall theory. The
thermodynamics of black holes in Rastall gravity have also been
studied \cite{18}.

To create new models of relativistic objects with unique properties,
Ovalle introduced the gravitational decoupling technique \cite{24}.
In many situations \cite{23}-\cite{22}, this technique has shown to
be a useful theoretical method for constructing potential star
distributions. Many authors \cite{25} used this procedure to
describe the inner region of self-gravitating compact structures
with both exotic as well as realistic fluids having anisotropic
distributions. Due to the extreme nonlinearity of the field
equations, there are relatively few physically sound analytical
solutions accessible, unless there are very particular constraints.
The division of the source energy-momentum into two halves is the
foundation of the gravitational decoupling extension. The first is
selected to produce a known solution of GR, while the second one
relates to an extra source that may contain any kind of charge, such
as gauge and tidal charges, or hairy fields related to gravity
outside of GR.

The gravitational decoupling technique, however, comes in two folds
viz the minimal geometric deformation (MGD) and the extended
geometric deformation (EGD). The fundamental difference between the
MGD and EGD is that the former transforms only the $g_{rr}$
component of the spacetime metric while the latter transforms both
the $g_{tt}$ and $g_{rr}$ metric coefficients of the spacetime.
Additionally, the MGD is limited in that it only applies when the
decoupled sources have a strictly gravitational interaction. This
implies that it is inapplicable in a scenario where there is energy
exchange between the decoupled sources. Using the deformation on
both temporal and radial metric functions, Casadio et al. \cite{26}
proposed an enhanced version of the MGD approach to address this
problem and produced a new solution for spherically symmetric
spacetime. However, since the conservation law breaks down in the
presence of matter, this extension is limited to studying vacuum
solutions. As such, this extended technique cannot be used to
discuss the intrinsic features and internal structure of
self-gravitating objects. Despite the said limitations, the MGD
approach has in recent time registered great success in obtaining
anisotropic solutions to different realistic compact stellar
configurations in GR \cite{27} as well as in various modified
theories \cite{28}, including the Rastall theory \cite{28a}. With
regards to black holes, we have used the MGD scheme to extend the
well known regular Hayward and Bardeen black hole solutions in
Rastall theory \cite{29}. Furthermore, Ovalle has used the MGD
decoupling to extend the Schwarzschild black hole in GR \cite{29a}.

By altering both metric potentials, Ovalle \cite{30} introduced the
innovative concept of EGD, which is applicable throughout spacetime
and is independent of the choice of matter distribution. The EGD has
since been exploited by numerous researchers to derive anisotropic
spherical solutions in a cascade of alternative theories. Contreras
and Bargueno applied the EGD scheme to vacuum BTZ solution in
$2+1$-dimensions \cite{31}. Sharif and Mughani \cite{32} used the
same scheme and extended the Tolman IV and Krori-Barua solutions,
respectively, to derive anisotropic solutions. Sharif and Majid
\cite{37a} found extended gravitationally decoupled solutions of the
Krori-Barua and Tolman IV ansatz in self-interacting Brans-Dicke
theory. Ovalle and his collaborators \cite{33} exploited this scheme
to obtain hairy black holes by extending the vacuum Schwarzschild
spacetime. Sharif and Majid \cite{37b} explored the effects of
charge on decoupled solutions in self-interacting Brans-Dicke (BD)
theory by taking Tolman IV and Krori-Barua solutions. The same
authors \cite{37} used the EGD method to extend the vacuum
Schwarzschild black hole in BD theory. Sharif and Naseer \cite{34}
investigated extended decoupled anisotropic solutions in
$f(\mathcal{R},T,\mathcal{R}_{ab}T^{ab})$ theory, using the EGD
technique both in the presence and absence of charge. The same
authors \cite{35} utilized this procedure to investigate its effects
on isotropization and complexity in $f(\mathcal{R},T)$ theory. They
\cite{36} also investigated the effect of charge on complexity and
isotropization of extended decoupled anisotropic stellar models in
the same theory.

In this work, we exploit the EGD approach to extend the vacuum
Schwarzschild black hole in Rastall theory. We obtain three
generalized solutions which we study in great detail and compare
with the earlier literature. The rest of this paper is aligned as
follows. Section \textbf{2} outlines the Rastall field equations for
a dual matter source and consequently defines some effective
parameters. We then apply the EGD technique to the field equations
(Section \textbf{3}). In Section \textbf{4}, we derive three
extended solutions and interpret these solutions via analysis of the
deformed metric potentials, the effective thermodynamic variables
and the energy conditions. Finally, we summarize our findings in a
well-articulated conclusion, in Section \textbf{5}.

\section{Rastall Field Equations}

The Rastall field equations are discriminated from the field
equations of GR through the Rastall parameter $\lambda$ which also
relates the covariant divergence of the Rastall stress-energy tensor
to that of the curvature scalar, $\mathcal{R}$. The field equations
for the Rastall theory can be formulated as
\begin{equation}\label{1}
G_{ab}+\frac{\lambda}{4}\mathcal{R}g_{ab}=\kappa T^R_{ab},
\end{equation}
where
\begin{equation}\label{2}
\nabla^b T^R_{ab}=\frac{\lambda}{4} g_{ab}\nabla^b\mathcal{R},
\end{equation}
describes the covariant divergence of the Rastall stress-energy
tensor $T^R_{ab}$. From Eq.\eqref{1}, $G_{ab}$ denotes the Einstein
tensor while $g_{ab}$ and $\kappa$ denote the metric tensor and
coupling constant, respectively.

The field equations \eqref{1} can be expressed in the alternate form
\begin{equation}\label{3}
G_{ab}=\kappa\bigg(T^R_{ab}-\frac{\lambda}{4(\lambda-1)}T^R
g_{ab}\bigg),
\end{equation}
by contracting them and using the resulting expression for the
curvature scalar $\mathcal{R}$. Equation \eqref{3} can be written as
\begin{equation}\label{4}
G_{ab}=\kappa T_{ab},
\end{equation}
if we define
\begin{equation}\label{4a}
T_{ab}=T^R_{ab}-\frac{\lambda}{4(\lambda-1)}T^R g_{ab}.
\end{equation}
This reorganization wherein the nonconforming terms of the Einstein
tensor are grouped, leading to the formation of an effective
stress-energy tensor, can be performed in any modified theory
irrespective of the state of conservation of their stress-energy
tensor. With this effective stress-energy tensor, the usual
conservation result $\nabla^b T_{ab}=0$ is regained. If we identify
$T_{ab}$ as the perfect fluid isotropic energy-momentum tensor of
GR, expressed as
\begin{equation}\label{4aa}
T_{ab}=(\rho+P)u_a u_b - Pg_{ab},
\end{equation}
with $~\rho,~P,u_a=\sqrt{g_{00}}\delta_a^0$ denoting the density,
isotropic pressure, and 4-velocity, respectively, then Eq.\eqref{4a}
relates the energy-momentum tensors of the GR and Rastall theories.
Further contracting Eq.\eqref{4a} gives the explicit relationship
\begin{equation}\label{4b}
(1-\lambda)T=T^R,
\end{equation}
between $T$ and $T^R$, the traces of the GR and Rastall
stress-energy tensors, respectively. By this relation, the Rastall
stress-energy tensor can be expressed as
\begin{equation}\label{4bb}
T^R_{ab}=T_{ab}-\frac{\lambda}{4}Tg_{ab}.
\end{equation}

In order to use the gravitational decoupling technique to extend a
known solution, we consider the field equations \eqref{1} with the
modification
\begin{equation}\label{4c}
G_{ab}+\frac{\lambda}{4}\mathcal{R}g_{ab}=\kappa T^{(total)}_{ab},
\end{equation}
where
\begin{equation}\label{4cc}
T^{(total)}_{ab}=T^R_{ab}+\beta\Theta_{ab}.
\end{equation}
Equation \eqref{4cc} shows that the total energy-momentum tensor
comprises a seed source $(T^R_{ab})$ to which an extra matter source
$(\Theta_{ab})$ is gravitationally coupled via the decoupling
parameter, $\beta$. This extra source may contain new fields of
scalars, vectors, and tensors and it is responsible for generating
anisotropy in the fluid. This total energy-momentum tensor must
(owing to its definition) satisfy the conservation equation
\begin{equation}\label{4d}
T^{a(total)}_{b;a}=0.
\end{equation}

We employ the following static spherically symmetric metric to
describe our spacetime geometry
\begin{equation}\label{5}
ds^2=e^{a(r)}dt^2-e^{b(r)}dr^2-r^2(d\theta^2+\sin^2\theta d\phi^2).
\end{equation}
This metric satisfies the field equations \eqref{1}, given by the
system below
\begin{align}\nonumber
\kappa\left[\rho-\frac{\lambda}{4}(\rho-3P)+\beta\Theta^0_0
\right]&=\frac{1}{r^2}+e^{-b}\bigg(\frac{b^\prime}{r}
-\frac{1}{r^2}\bigg)+\frac{\lambda e^{-b}}{4}
\bigg(a^{\prime\prime}+\frac{a^\prime(a^\prime
-b^\prime)}{2}\bigg)\\\label{6} &+\frac{\lambda
e^{-b}}{4}\bigg(\frac{2(a^\prime -b^\prime)}{r}
+\frac{2}{r^2}\bigg)-\frac{\lambda}{2r^2},
\end{align}
\begin{align}\nonumber
\kappa\left[P+\frac{\lambda}{4}(\rho-3P)-\beta\Theta^1_1\right]&=
-\frac{1}{r^2}+e^{-b}\bigg(\frac{a^\prime}{r}+\frac{1}{r^2}\bigg)
-\frac{\lambda e^{-b}}{4}\bigg(a^{\prime\prime}
+\frac{a^\prime(a^\prime-b^\prime)}{2}\bigg)\\\label{7}
&-\frac{\lambda e^{-b}}{4}\bigg(\frac{2(a^\prime -b^\prime)}{r}
+\frac{2}{r^2}\bigg)+\frac{\lambda}{2r^2},
\end{align}
\begin{align}\nonumber
\kappa\left[P+\frac{\lambda}{4}(\rho-3P)-\beta\Theta^2_2\right]&=e^{-b}
\bigg(\frac{a^{\prime\prime}}{2}+\frac{(a^\prime)^2}{4}
-\frac{a^\prime b^\prime}{4}+\frac{a^\prime}{2r}
-\frac{b^\prime}{2r}\bigg)+\frac{\lambda}{2r^2}\\\label{8}
&-\frac{\lambda e^{-b}}{4}\bigg(a^{\prime\prime}
+\frac{a^\prime(a^\prime-b^\prime)}{2}
+\frac{2(a^\prime-b^\prime)}{r}+\frac{2}{r^2}\bigg).
\end{align}
The conservation equation \eqref{4d} with respect to the system
above is given as
\begin{equation}\label{8a}
\frac{dP(r)}{dr}+\frac{a^\prime(r)}{2}(\rho+P)
+\frac{2\beta}{r}(\Theta^2_2-\Theta^1_1)+\frac{\beta
a^\prime(r)}{2}(\Theta^0_0-\Theta^1_1)-\beta\bigg(\Theta^1_1(r)\bigg)^\prime=0.
\end{equation}
This system comprises three nonlinear ordinary differential
equations in the seven unknowns
$a(r),~b(r),~\rho(r)~,P(r),~\Theta_0^0,~\Theta_1^1,~\Theta^2_2$, and
$^\prime=\frac{d}{dr}$. It is from this system that the following
effective parameters are identified
\begin{equation}\label{9}
\rho^{eff}=\rho+\beta\Theta^0_0,\quad
P_r^{eff}=P-\beta\Theta^1_1,\quad P_t^{eff}=P-\beta\Theta^2_2.
\end{equation}
These effective parameters imply an anisotropy induced by the extra
source $\Lambda_{ab}$, given by
\begin{equation}\label{10}
\Delta^{eff}=P_t^{eff}-P_r^{eff}=\beta\big(\Theta^1_1-\Theta^2_2\big),
\end{equation}
which vanishes only in the event $\Theta^1_1=\Theta^2_2$.

\section{Extended Geometric Deformation Technique}

In a bid to solve the system \eqref{6}-\eqref{8}, we employ the EGD
technique which deforms both the temporal and radial metric
coefficients by means of some appropriate linear transformations.
Using these transformations, the system splits into two sets, the
first of which corresponds to a perfect fluid matter distribution
($\beta=0$). The second set entails the additional source
$\Lambda_{ab}$ and depicts a quasi-Einstein system. Proceeding, we
consider a known ideal fluid solution to the field equations,
described by the metric
\begin{equation}\label{11}
ds^2=e^{\sigma(r)}dt^2-\frac{1}{\eta(r)}dr^2-r^2(d\theta^2+\sin^2\theta
d\phi^2),
\end{equation}
with
\begin{equation}\label{12}
\eta(r)=1-\frac{2m(r)}{r},
\end{equation}
where $m$ represents the Misner-Sharp mass. The linear
transformations that characterize the geometric deformation are
given by
\begin{equation}\label{13}
\sigma(r)\mapsto a(r)=\sigma(r)+\beta f_1(r),\quad\eta(r)\mapsto
e^{-b(r)}=\eta(r)+\beta f_2(r),
\end{equation}
where $f_1(r)$ and $f_2(r)$ denote the deformations applied to the
temporal and radial metric components, respectively. Substituting
the transformations \eqref{13} into the field equations, we obtain
the first set as
\begin{align}\nonumber
\kappa\left[\rho-\frac{\lambda}{4}(\rho-3P)\right]&=\eta\bigg(\frac{\lambda\sigma^{\prime\prime}}{4}
-\frac{1}{r^2}+\frac{\lambda(\sigma^\prime)^2}{8}+\frac{\lambda\sigma^\prime}{2r}
+\frac{\lambda}{2r^2}\bigg)\\\label{14}&+\eta^\prime\bigg(\frac{\lambda}{2r}
+\frac{\lambda\sigma^\prime}{8}-\frac{1}{r}\bigg)-\frac{\lambda}{2r^2}+\frac{1}{r^2},
\end{align}
\begin{align}\nonumber
\kappa\left[P+\frac{\lambda}{4}(\rho-3P)\right]&=\eta\bigg(\frac{\sigma^\prime}{r}
-\frac{\lambda\sigma^{\prime\prime}}{4}+\frac{1}{r^2}
-\frac{\lambda(\sigma^\prime)^2}{8}-\frac{\lambda\sigma^\prime}{2r}
-\frac{\lambda}{2r^2}\bigg)\\\label{15}
&-\eta^\prime\bigg(\frac{\lambda\sigma^\prime}{8}
+\frac{\lambda}{2r}\bigg)+\frac{\lambda}{2r^2}-\frac{1}{r^2},
\end{align}
\begin{align}\nonumber
\kappa\left[P+\frac{\lambda}{4}(\rho-3P)\right]&=\eta\bigg(\frac{\sigma^{\prime\prime}}{2}
+\frac{(\sigma^\prime)^2}{4}+\frac{\sigma^\prime}{2r}
-\frac{\lambda\sigma^{\prime\prime}}{4}
-\frac{\lambda(\sigma^\prime)^2}{8}-\frac{\lambda\sigma^\prime}{2r}
-\frac{\lambda}{2r^2}\bigg)\\\label{16}&+\eta^\prime\bigg(\frac{\sigma^\prime}{4}
+\frac{1}{2r}-\frac{\lambda\sigma^\prime}{8}-\frac{\lambda}{2r}\bigg)+\frac{\lambda}{2r^2},
\end{align}
associated to the conservation equation
\begin{equation}\label{16a}
\frac{dP(r)}{dr}+\frac{a^\prime(r)}{2}(\rho+P)=0.
\end{equation}
By adopting any known spherically symmetric solution for the metric
potentials ($\sigma$ and $\eta$) and expressing the density and
pressure in terms of these metric potentials, the system
Eqs.\eqref{14}-\eqref{16} can be solved.

The second set is given by the following system
\begin{align}\nonumber
\kappa(\Theta^0_0)&=\frac{\lambda}{4}\bigg[f_2\bigg(
\sigma^{\prime\prime}+\frac{(\sigma^\prime)^2}{2}
+\frac{2\sigma^\prime}{r}\bigg)+f_2^\prime\bigg(
\frac{\sigma^\prime}{2}+\frac{2}{r}\bigg)+\eta
f_1^{\prime\prime}+\eta\sigma^\prime
f_1^\prime+\frac{\eta\beta(f_1^\prime)^2}{2}\\\label{17}
&+\frac{\eta^\prime f_1^\prime}{2}+\frac{2\eta
f_1^\prime}{r}\bigg]-\frac{f_2^\prime}{r}-\frac{f_2}{r^2},
\end{align}
\begin{align}\nonumber
\kappa(\Theta^1_1)&=\frac{\lambda}{4}\bigg[f_2\bigg(
\sigma^{\prime\prime}+\frac{(\sigma^\prime)^2}{2}
+\frac{2\sigma^\prime}{r}\bigg)+f_2^\prime\bigg(
\frac{\sigma^\prime}{2}+\frac{2}{r}\bigg)+\eta
f_1^{\prime\prime}+\eta\sigma^\prime
f_1^\prime+\frac{\eta\beta(f_1^\prime)^2}{2}\\\label{18}
&+\frac{\eta^\prime f_1^\prime}{2}+\frac{2\eta
f_1^\prime}{r}\bigg]-f_2\bigg(\frac{\sigma^\prime}{r}
+\frac{1}{r^2}\bigg)-\frac{\eta f_1^\prime}{r},
\end{align}
\begin{align}\nonumber
\kappa(\Theta^2_2)&=\frac{\lambda}{4}\bigg[f_2\bigg(
\sigma^{\prime\prime}+\frac{(\sigma^\prime)^2}{2}
+\frac{2\sigma^\prime}{r}\bigg)+f_2^\prime\bigg(
\frac{\sigma^\prime}{2}+\frac{2}{r}\bigg)+\eta
f_1^{\prime\prime}+\eta\sigma^\prime
f_1^\prime+\frac{\eta\beta(f_1^\prime)^2}{2}\\\nonumber&+\frac{\eta^\prime
f_1^\prime}{2}+\frac{2\eta f_1^\prime}{r}\bigg]
-f_2\bigg(\frac{\sigma^{\prime\prime}}{2}
+\frac{(\sigma^\prime)^2}{4}+\frac{\sigma^\prime}{2r}\bigg)
-f_2^\prime\bigg(\frac{\sigma^\prime}{4}+\frac{1}{2r}\bigg)
-\frac{\eta^\prime f_1^\prime}{4}\\\label{19}&-\eta\bigg(
\frac{f_1^{\prime\prime}}{2}+\frac{\beta(f_1^\prime)^2}{4}
+\frac{\eta^\prime f_1^\prime}{2}+\frac{f_1^\prime}{2r}\bigg),
\end{align}
and conserves according to the equation
\begin{equation}\label{19a}
\frac{2\beta}{r}(\Theta^2_2-\Theta^1_1)+\frac{\beta
a^\prime(r)}{2}(\Theta^0_0-\Theta^1_1)-\beta\bigg(\Theta^1_1(r)\bigg)^\prime=0.
\end{equation}
The system \eqref{17}-\eqref{19} above is solved by imposing two
constraints (as there are five unknowns in three equations). The
first of these constraints will be applied to the metric potentials
while the second will be imposed on the extra source via a linear
equation of state (EoS). A solution to the field equations
\eqref{6}-\eqref{8} is thus found by the superposition principle,
via a combination of the solutions of the two systems above, such as
given by Eq.\eqref{9}.

\section{Extended Schwarzschild Solutions}

Here, we develop from the results of the previous section and obtain
the deformation functions $f_1(r)$ and $f_2(r)$ for the
Schwarzschild vacuum solution, given by
\begin{equation}\label{20}
ds^2=\bigg(1-\frac{2M}{r}\bigg)dt^2-\bigg(1-\frac{2M}{r}\bigg)
^{-1}dr^2-r^2d\theta^2-r^2\sin^2\theta d\phi^2,
\end{equation}
where $M$ denotes the Schwarzschild mass. Due to the presence of a
vacuum, we have $\rho=P=0$, thus redefining the effective parameters
in Eq.\eqref{9}. The transformations in Eq.\eqref{13} deform the
Schwarzschild metric as follows
\begin{equation}\label{21}
ds^2=\bigg(1-\frac{2M}{r}\bigg)e^{\beta
f_1(r)}dt^2-\frac{dr^2}{\bigg(1-\frac{2M}{r}+\beta
f_2(r)\bigg)}-r^2\bigg(d\theta^2+\sin^2\theta d\phi^2\bigg),
\end{equation}
yielding the extended Schwarzschild solution. From the Schwarzschild
metric \eqref{20}, we observe the overlap of the Killing ($r_H$) and
the causal horizons ($r_h$), at the surface $r=2M$. These horizons
are determined by $e^a=0$ and $e^{-b}=0$, respectively \cite{29}. It
can also be observed that there lies a singularity at $r=0$, behind
the Killing and causal horizons. By the coincidence of these, a
prerequisite is obtained for the EGD metric \eqref{21} to depict a
well-defined black hole. This coincidence implies that $e^a=e^{-b}$,
which yields the first constraint
\begin{equation}\label{22}
a=-b.
\end{equation}
Using this constraint in the transformation equations \eqref{13}, we
obtain the fundamental relationship between the deformation
functions, given by
\begin{equation}\label{23}
f_2(r)=\frac{(r-2M)(e^{\beta f_1(r)}-1)}{\beta r}.
\end{equation}
The second constraint needed to evaluate the deformation functions
$f_1$ and $f_2$ is given via the following linear EoS \cite{29}
\begin{equation}\label{24}
\Theta_0^0+\alpha_1\Theta_1^1=\alpha_2\Theta^2_2,
\end{equation}
with $\alpha_1$ and $\alpha_2$ are arbitrary constants.

Following are three extensions of the Schwarzschild black hole, each
of which is obtained using a particular case of the EoS \eqref{24}.

\subsection{Model I: Traceless $\Theta_b^a$}

Since $\Theta^2_2=\Theta^3_3$, it is reasonable to assume that the
extra source has a trace-free energy-momentum tensor when
$\alpha_1=1$ and $\alpha_2=-2$ in \eqref{24}, i.e.,
\begin{equation}\label{25}
\Theta_0^0+\Theta_1^1=-2\Theta^2_2.
\end{equation}
Utilizing the system \eqref{17}-\eqref{19}, Eq.\eqref{25} becomes
\begin{align}\nonumber
&-f_2\bigg(\frac{\sigma^\prime}{r}+\frac{2}{r^2}\bigg)
-\frac{f_2^\prime}{r}-\frac{\eta f_1^\prime}{r}
+\lambda\bigg[f_2\bigg(\sigma^{\prime\prime}
+\frac{\sigma^{\prime^2}}{2}+\frac{2\sigma^\prime}{r}\bigg)
+f_2^\prime\bigg(\frac{\sigma^\prime}{2}+\frac{2}{r}\bigg)+\eta
f_1^{\prime\prime}\\\nonumber&+\eta\sigma^\prime f_1^\prime
+\frac{\eta\beta f_1^{\prime^2}}{2}+\frac{\eta^\prime
f_1^\prime}{2}+\frac{2\eta f_1^\prime}{r}\bigg]
-f_2\bigg(\sigma^{\prime\prime}+\frac{\sigma^{\prime^2}}{2}
+\frac{\sigma^\prime}{r}\bigg)-f_2^\prime\bigg(\frac{\sigma^\prime}{2}
+\frac{1}{r}\bigg)-\frac{\eta^\prime f_1^\prime}{2}\\\label{26}&-
\eta\bigg(f_1^{\prime\prime}+\frac{\beta
f_1^{\prime^2}}{2}+\sigma^\prime
f_1^\prime+\frac{f_1^\prime}{r}\bigg)=0.
\end{align}
Using the equation above together with the relation given by
Eq.\eqref{23}, we obtain numerical approximations for the
deformation functions $f_1$ and $f_2$. Inserting these
approximations of $f_1$ and $f_2$ in the EGD metric \eqref{21}, we
obtain our required solutions.

We present the graphs of the distorted metric, from which we analyze
the asymptotic flatness of the obtained spacetime. A spacetime is
termed as asymptotically flat if the metric potentials tend to $1$
when the radial coordinate is taken to be sufficiently large. In
such a spacetime, the effect of a gravitational field decreases and
becomes unnoticeable after traveling a large distance from a
gravitating body, thus the space outside this region looks almost
flat. The Rastall and decoupling parameters $\lambda=0.01$ (solid
lines), $0.02$ (dashed lines), and $\beta=-0.1$ (blue), $-0.102$
(brown), $-0.104$ (green), $-0.106$ (red), $-0.108$ (black) were
used consistently in all calculations. We take $M=1$ as the mass,
and this consideration contains an area that an observer can reach.
\begin{figure}\center
\epsfig{file=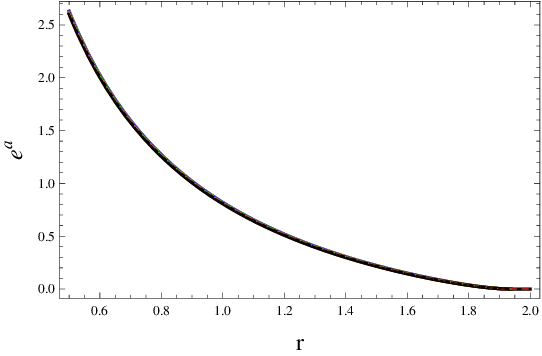,width=0.475\linewidth}
\epsfig{file=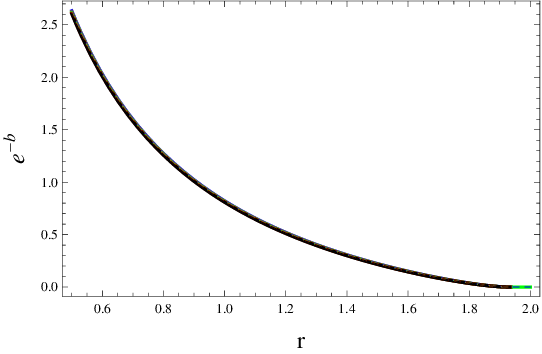,width=0.475\linewidth}\caption{Graphs of deformed
metric coefficients $e^{a}$ and $e^{-b}$ against $r$ model I.}
\end{figure}
\begin{figure}\center
\epsfig{file=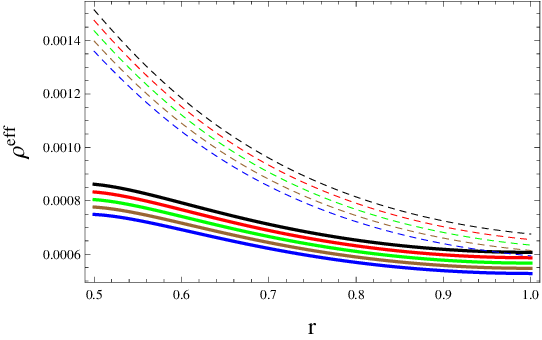,width=0.475\linewidth}
\epsfig{file=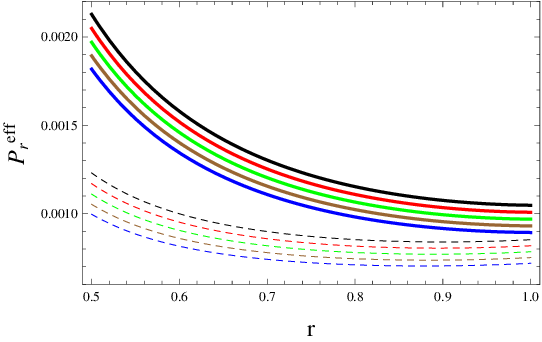,width=0.475\linewidth}
\epsfig{file=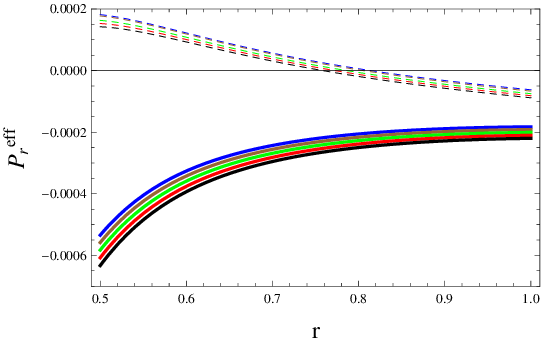,width=0.475\linewidth}\caption{Graphs of
$\rho^{eff},P_r^{eff},P_t^{eff}$ against $r$ for model I.}
\end{figure}

The plots of the deformed metric coefficients displayed in Figure
\textbf{1} show that the resulting spacetime fails to preserve
asymptotic flatness. We further plot the effective parameters
(Figure \textbf{2}) to describe the nature of our solution. The
density turns out to be positive, while a positive radial pressure
is obtained. Positive radial pressure means that there is an outward
pressure opposing the inward pull of gravity. This is not a typical
situation for a black hole because the strong gravitational forces
inside a black hole are typically associated with very high inward
pressure. However, in theoretical models involving unusual forms of
matter or energy, like negative mass or energy, it is possible that
there could be a positive radial pressure acting outward within the
black hole. It is worthy to mention that a positive energy density
could only be obtained for negative values of the decoupling
parameter, $\beta$. With respect to the Rastall parameter
($\lambda$), both the energy density and tangential pressure vary
directly, while the radial pressure varies inversely. With regards
to the decoupling parameter, the energy density and radial pressure
vary inversely, while the tangential pressure varies directly.
\begin{figure}\center
\epsfig{file=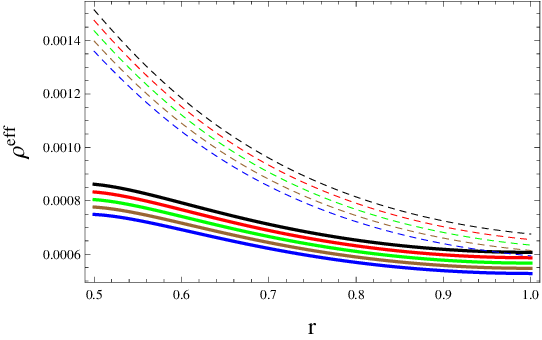,width=0.475\linewidth}
\epsfig{file=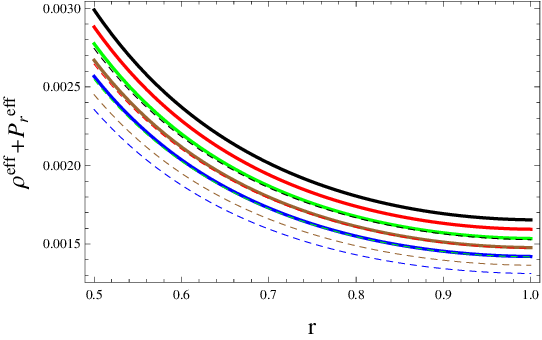,width=0.475\linewidth}
\epsfig{file=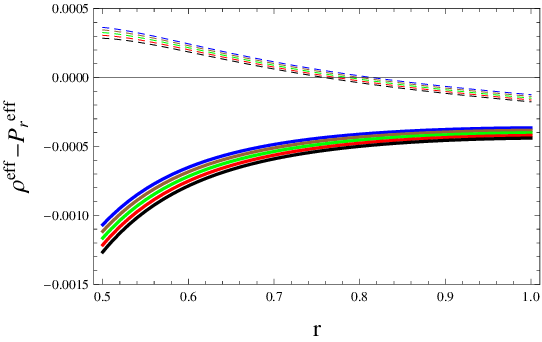,width=0.475\linewidth}
\epsfig{file=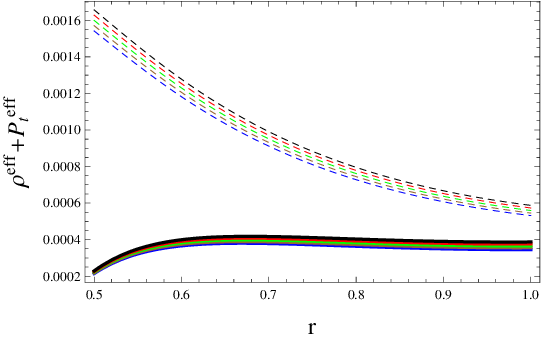,width=0.475\linewidth}
\epsfig{file=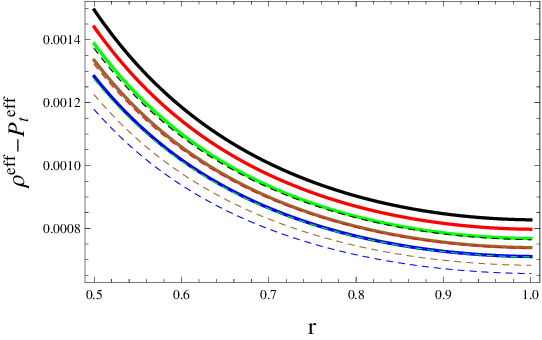,width=0.475\linewidth}
\epsfig{file=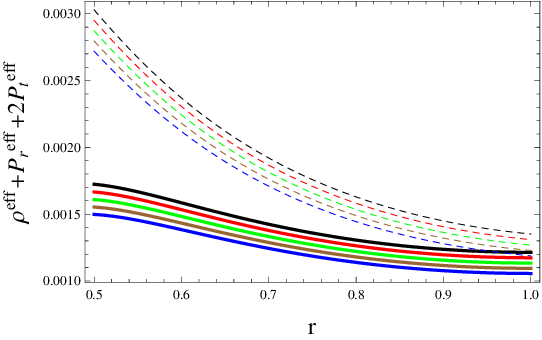,width=0.475\linewidth} \caption{Graphs of energy
bounds against $r$ for model I.}
\end{figure}

It is interesting to observe that for both values of the Rastall
parameter, the radial pressure is maximum around the core and
monotonically decreases towards the event horizon. However, the
lower value of the Rastall parameter ($\lambda=0.01$) induces a much
higher pressure at the core. The difference in the radial pressures
corresponding to the two Rastall parameters is thus more significant
around the core while vanishing towards the surface. Similarly, the
energy density is maximum at the core (for both values of the
Rastall parameter) and decreases monotonically towards the surface.
Contrary to the radial pressure, the higher value of the Rastall
parameter ($\lambda=0.02$) induces a more dense core. As with the
radial pressure, the difference in the densities (with respect to
the two values of the Rastall parameter used) is greater around the
core while disappearing towards the surface.

Finally, we investigate the adherence of the thermodynamic variables
to the following energy conditions
\begin{align}\nonumber
\rho^{eff}\geq 0,\quad\rho^{eff}+P_r^{eff}\geq 0,\\\label{27}
\rho^{eff}-P_r^{eff}\geq 0,\quad\rho^{eff}-P_t^{eff}\geq
0,\\\nonumber \rho^{eff}+P_t^{eff}\geq
0,\quad\rho^{eff}+P_r^{eff}+2P_t^{eff}\geq 0.
\end{align}
These conditions, if satisfied, imply that the matter is ordinary.
Failure to satisfy these energy conditions imply that the matter is
exotic. The plots of the energy conditions in Figure \textbf{3}
portray that the matter source is exotic, due to the violation of a
dominant energy condition.

\subsection{Model II: A barotropic EoS}

The source $\Theta_{ab}$ is termed as a polytropic fluid in the
event if it satisfies the following EoS \cite{29}
\begin{equation}\label{28}
P_r^{eff}-\varrho\bigg(\rho^{eff}\bigg)^\Gamma,
\end{equation}
where $\varrho>0$ contains parametric information about the
temperature and $\Gamma=1+\frac{1}{n},~n$ is the polytropic index.
Using the appropriate substitutions and considering the unique case
$\Gamma=1$, the equation above simplifies to
\begin{equation}\label{29}
\varrho(\Theta^0_0)+\Theta_1^1=0,
\end{equation}
denoting a barotropic EoS \cite{29}. Equation \eqref{29} can be
identified as a particular case of the EoS \eqref{24}, with
$\alpha_1=\frac{1}{\varrho}$ and $\alpha_2=0$. Using Eqs.\eqref{17}
and \eqref{18}, this equation gives
\begin{align}\nonumber
&-\varrho\bigg(\frac{f_2^\prime}{r}+\frac{f_2}{r^2}\bigg)
-f_2\bigg(\frac{\sigma^\prime}{r}+\frac{1}{r^2}\bigg)-\frac{\eta
f_1^\prime}{r}+\frac{\lambda(\varrho+1)}{4}\bigg[f_2\bigg(\sigma^{\prime\prime}
+\frac{\sigma^{\prime^2}}{2}+\frac{2\sigma^\prime}{r}\bigg)\\\label{30}&
+f_2^\prime\bigg(\frac{\sigma^\prime}{2}+\frac{2}{r}\bigg) +\eta
f_1^{\prime\prime}+\eta\sigma^\prime f_1^\prime +\frac{\eta\beta
f_1^{\prime^2}}{2}+\frac{\eta^\prime f_1^\prime}{2}+\frac{2\eta
f_1^\prime}{r}\bigg]=0.
\end{align}
Using this equation together with Eq.\eqref{23}, we obtain numerical
approximations of the functions $f_1$ and $f_2$ which are then
applied to the EGD metric \eqref{21} to obtain the required
solution.
\begin{figure}\center
\epsfig{file=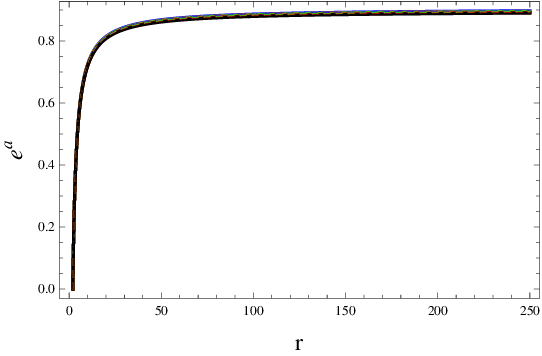,width=0.475\linewidth}
\epsfig{file=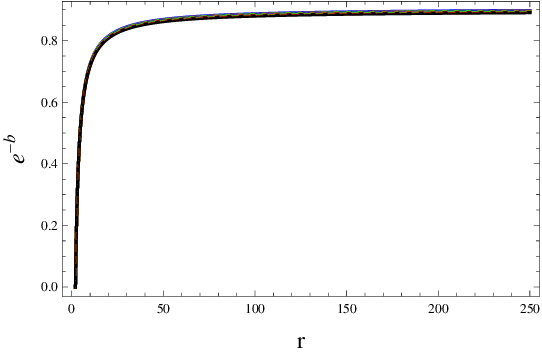,width=0.475\linewidth}\caption{Graphs of deformed
metric coefficients $e^{a}$ and $e^{-b}$ against $r$ for Model II.}
\end{figure}

We plot the metric potentials in Figure \textbf{4} from which it is
seen that the resulting spacetime is almost asymptotically flat, as
the metric potentials approach $0.9$ (approximately 1) as r
increases arbitrarily. The effective thermodynamic variables plotted
in Figure \textbf{5}, show a positive density, a negative radial
pressure and a positive tangential pressure. An inward pressure is
implied by a negative radial pressure, which strengthens the black
hole's gravitational attraction. This idea is more consistent with
the known information about black holes, where matter collapses to a
singularity due to extremely high gravitational forces. In
theoretical discussions, negative radial pressure is frequently used
to explain phenomena such as the universe's accelerated expansion in
theories including dark energy with negative pressure. It is
observed that the increment in the Rastall parameter registers no
significance in the outputs of the energy density and radial
pressure. This increment, however, induces a higher tangential
pressure. Both energy density and tangential pressure are maximum at
the core and decrease monotonically towards the surface. To the
contrary, the radial pressure is minimum at the core and increases
monotonically towards the surface. With respect to the decoupling
parameter, the density and tangential pressure vary inversely while
the radial pressure vary directly. Finally, the analysis of the
energy conditions in Figure \textbf{6} reveals an exotic source due
to the violation of one of the dominant energy conditions.
\begin{figure}\center
\epsfig{file=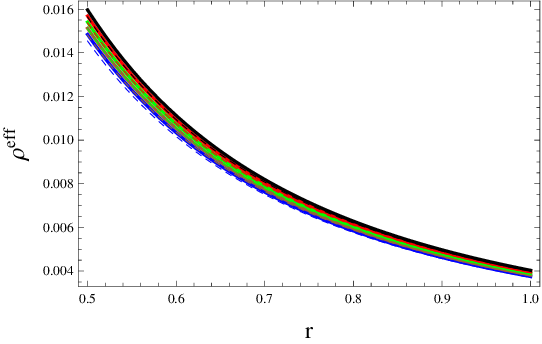,width=0.475\linewidth}
\epsfig{file=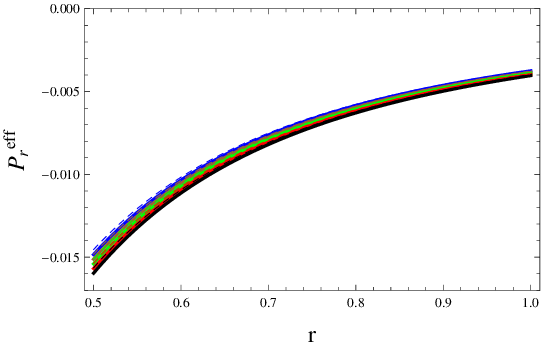,width=0.475\linewidth}
\epsfig{file=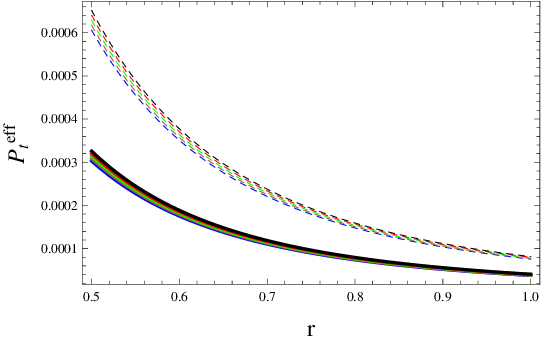,width=0.475\linewidth}\caption{Graphs of
$\rho^{eff},P_r^{eff},P_t^{eff}$ against $r$ for model II.}
\end{figure}
\begin{figure}\center
\epsfig{file=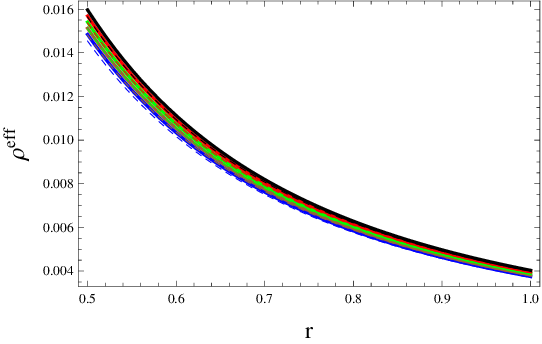,width=0.475\linewidth}
\epsfig{file=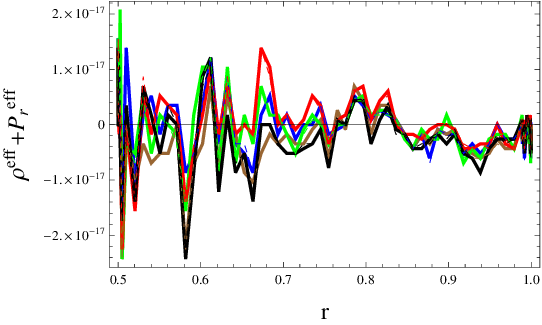,width=0.475\linewidth}
\epsfig{file=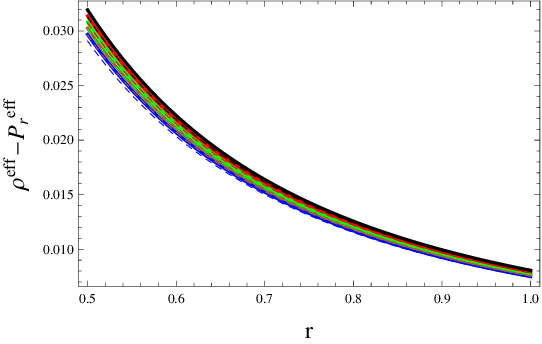,width=0.475\linewidth}
\epsfig{file=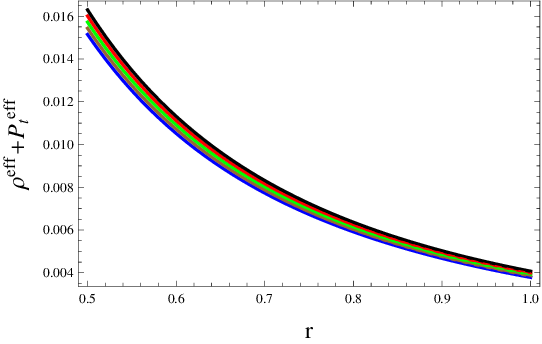,width=0.475\linewidth}
\epsfig{file=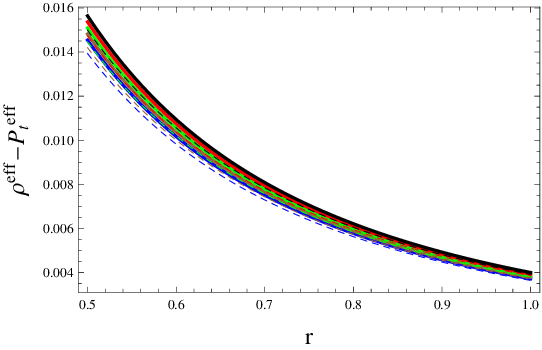,width=0.475\linewidth}
\epsfig{file=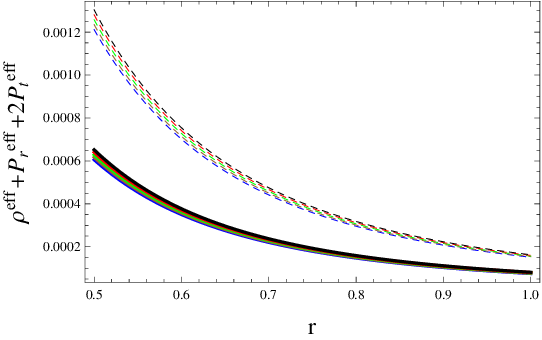,width=0.475\linewidth} \caption{Graphs of energy
bounds against $r$ for model II.}
\end{figure}

\subsection{Model III: A Particular Case}

We consider a unique case of the EoS \eqref{24} with $\alpha_1=1.4$
and $\alpha_2=-3$ as in \cite{29}, thus giving the linear equation
\begin{equation}\label{31}
\Theta_0^0+1.4\Theta^1_1+3\Theta^2_2=0.
\end{equation}
This equation leads to
\begin{align}\nonumber
&-3f_2\bigg(\frac{\sigma^{\prime\prime}}{2}+
\frac{\sigma^{\prime^2}}{4}+\frac{\sigma^\prime}{2r}
\bigg)-\frac{3\sigma^\prime
f_1^\prime}{4}-3\eta\bigg(\frac{f_1^{\prime\prime}}{2}+\frac{\beta
f_1^{\prime^2}}{4}+\frac{\sigma^\prime
f_1^\prime}{2}+\frac{f_1^\prime}{2r}\bigg)\\\nonumber&
+\frac{\lambda}{4}\bigg[f_2\bigg(\sigma^{\prime\prime}
+\frac{\sigma^{\prime^2}}{2}+\frac{2\sigma^\prime}{r}\bigg)
+f_2^\prime\bigg(\frac{\sigma^\prime}{2}+\frac{2}{r}\bigg) +\eta
f_1^{\prime\prime}+\eta\sigma^\prime f_1^\prime +\frac{\eta\beta
f_1^{\prime^2}}{2}+\frac{\eta^\prime f_1^\prime}{2}
\\\label{32}&+\frac{2\eta f_1^\prime}{r}\bigg]-\frac{f_2^\prime}{r}
-\frac{f_2}{r^2}-1.4f_2\bigg(\frac{\sigma^\prime}{r}+\frac{1}{r^2}\bigg)
-\frac{1.4\eta f_1^\prime}{r}
-3f_2^\prime\bigg(\frac{\sigma^\prime}{4} +\frac{1}{2r}\bigg)=0.
\end{align}
As with the cases of the previously obtained models, we use the
equation above together with the relation given by Eq.\eqref{23} to
obtain the functions $f_1$ and $f_2$. The metric for the spacetime
describing the obtained black hole can then be evaluated by using
these functions in the EGD metric \eqref{21}. Figure \textbf{7}
shows the plots of the metric potentials which show that the
resulting spacetime is not asymptotically flat. The effective
thermodynamic variables are shown in Figure \textbf{8}. A positive
energy density and a negative tangential pressure are observed, both
of which vary directly with the Rastall parameter. The radial
pressure turns out to be positive and varies inversely with the
Rastall parameter. The decoupling parameter, however, varies
inversely with the energy density and radial pressure, while
exhibiting direct proportionality to the tangential pressure. It can
be observed for this model that the variation of the Rastall
parameter registers a significant difference in the output of all
three thermodynamic variables. Lastly, we plot the energy conditions
which again show a violation of some dominant conditions (Figure
\textbf{9}).
\begin{figure}\center
\epsfig{file=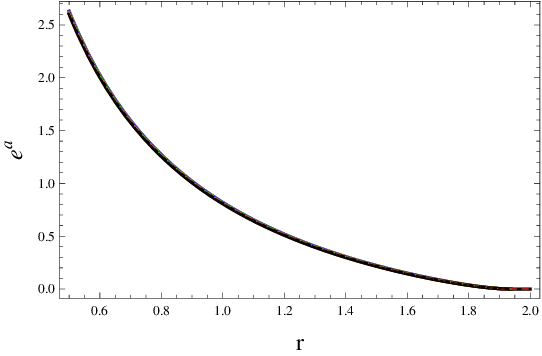,width=0.475\linewidth}
\epsfig{file=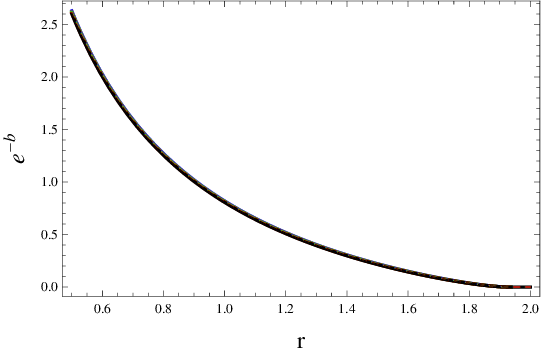,width=0.475\linewidth} \caption{Graphs of deformed
metric coefficients $e^{a}$ and $e^{-b}$ against $r$ for model III.}
\end{figure}
\begin{figure}\center
\epsfig{file=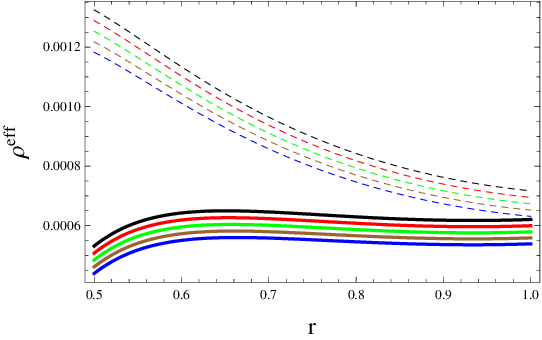,width=0.475\linewidth}
\epsfig{file=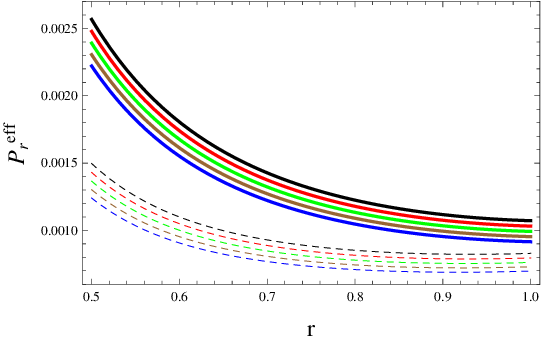,width=0.475\linewidth}
\epsfig{file=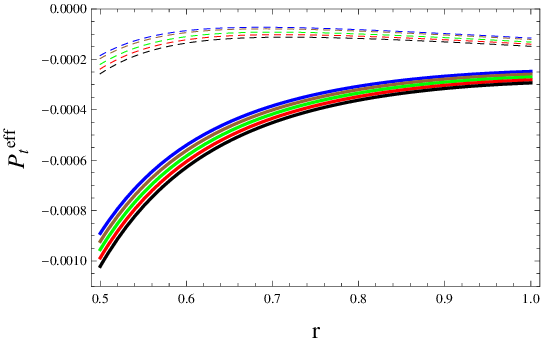,width=0.475\linewidth}\caption{Graphs of
$\rho^{eff},P_r^{eff},P_t^{eff}$ against $r$ for model III.}
\end{figure}
\begin{figure}\center
\epsfig{file=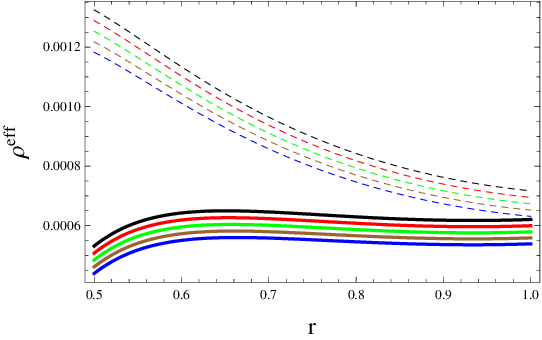,width=0.475\linewidth}
\epsfig{file=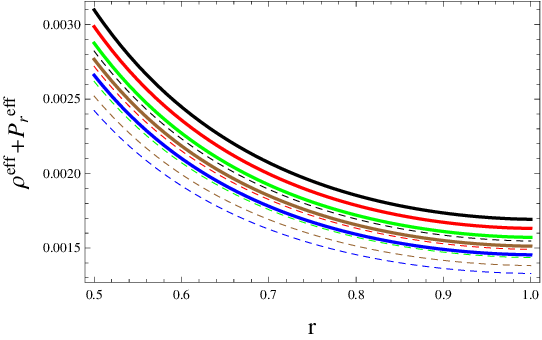,width=0.475\linewidth}
\epsfig{file=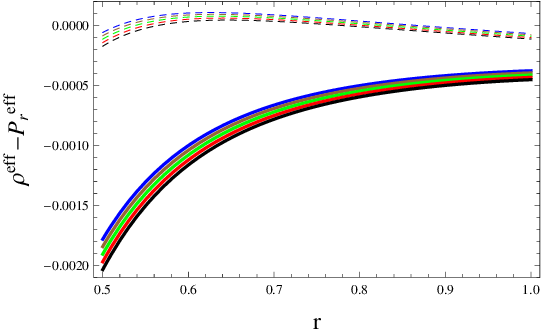,width=0.475\linewidth}
\epsfig{file=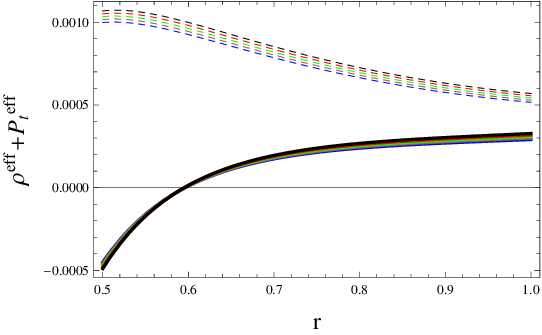,width=0.475\linewidth}
\epsfig{file=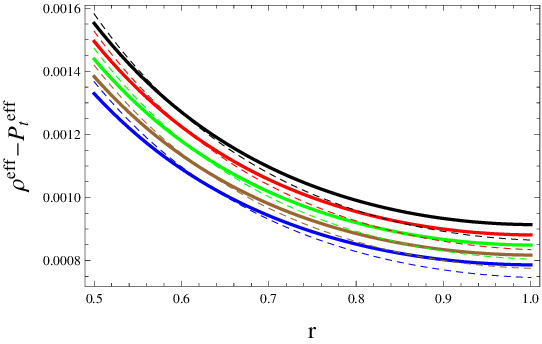,width=0.475\linewidth}
\epsfig{file=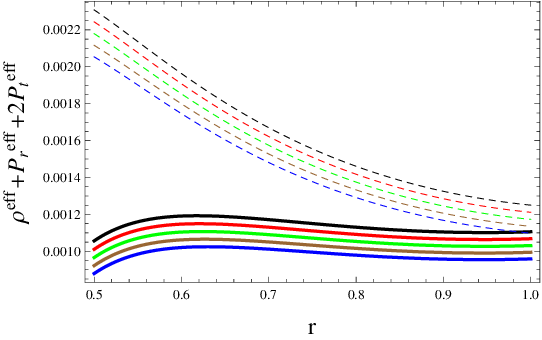,width=0.475\linewidth} \caption{Graphs of energy
bounds against $r$ for $\lambda=0.01$ (solid), $0.02$ (dashed),
$\beta=-0.1$ (blue), $-0.102$ (brown), $-0.104$ (green), $-0.106$
(red) and $-0.108$ (black) for model III.}
\end{figure}

\section{Conclusions}

This paper focuses on exploring the EGD method to extend the known
Schwarzschild black hole solution within the framework of the
Rastall theory of gravity. The field equations for an extra matter
source gravitationally coupled to a seed source of matter are
explicitly formulated. Under the EGD strategy, no restrictions are
made with regards to the nature of the interaction between the
coupled matter sources. These field equations are extensively
decoupled, thus generating two new systems which correspond to the
seed and extra source, respectively. This decoupling has been done
via linear transformations which alters both metric coefficients.

Due to the presence of a vacuum (which characterizes the seed
source), the obtained models are explicitly described by the extra
source, $\Theta_{ab}$. We have thus obtained the extended models by
imposing two appropriate constraints on the metric functions and on
this extra source. The first constraint ($a=-b$) which is imposed on
the metric functions ensures that the Killing and causal horizons
overlap, which is a necessary prerequisite for a well defined black
hole model. The second constraint is given by the linear EoS,
$\Theta_0^0+\alpha_1\Theta_1^1=\alpha_2\Theta^2_2$, through which
the extended models are obtained. Three models corresponding to
three cases of the mentioned EoS are thus obtained. The effect of
the Rastall and decoupling parameters, respectively, have been
extensively investigated for all the obtained models. To this
effect, we have adopted the values $\lambda=0.01,0.02$ and
$\beta=-0.1,-0.102,-0.104,-0.106,-0.108$. It is found that
increasing the Rastall parameter induces a denser core in models I
and III, while no effect is registered in model II. With regard to
the radial pressure, the effect of this increment marks a lower
radial pressure.

For all the obtained models, the energy density is found to be
positive, as required. A negative radial pressure (which is typical
for black holes) is only obtained for the second model given by a
barotropic EoS \eqref{31}. It is worthy to mention that all the
three cases considered in this work are also considered in GR
\cite{29} and BD theory \cite{37}. However, with regards to the
behavior of the thermodynamic variables, only the results of model
II are in line with the results of their counterparts in
\cite{29a,37}. Through analysis of the metric potentials, it is
found that only the second model tends to an asymptotically flat
spacetime. In GR \cite{29a}, the model obtained from the traceless
extra source (corresponding to our model I) failed to preserve
asymptotic flatness. In BD theory \cite{37}, however, it was found
that only the model generated using a barotropic EoS (corresponding
to our model II) failed to preserve asymptotic flatness.

It is found for all obtained models that the dominant energy
conditions are not satisfied, implying that the additional source is
exotic in all cases. This result was also obtained in GR \cite{29},
where it was concluded that the dominant energy conditions are
unsatisfiable using the EoS \eqref{24}. The metric potential
(Figures \textbf{1}, \textbf{4} and \textbf{7}) reveal that all the
obtained models have Killing and causal horizons coinciding at
$r=2M$, where ($M=1$). Another observation directed at the EGD
metric \eqref{21}, reveals that all the extended models have a
singularity at $r=0$. Thus putting together these two aforementioned
observations drive us to the conclusion that all the obtained models
have a singularity at $r=0$ hidden behind the horizon at $r=2M$.
This result is common to \cite{29a} and \cite{37}.\\\\
\textbf{Data Availability Statement:} No data was used for the
research described in this paper.

\end{document}